\documentstyle[a4,12pt]{article}
\newcommand{\mw}[1]{\langle#1\rangle} 
\newcommand{\rn}{\rho^n}
\newcommand{\rs}{\rho^s}

\newcommand{\dta}{\delta T_A}
\newcommand{\dtb}{\delta T_B}
\newcommand{\DTi}{\Delta T_i}
\newcommand{\dts}{\pdf{T}\sigma}
\newcommand{\delmu}{\Delta \mu_i}
\newcommand{\delsig}{\Delta \sigma_i}
\newcommand{\Rp}{\dot{R}}
\newcommand{\Rpinf}{\dot{R}_\infty}
\newcommand{\staAB}{\stackrel{\scriptstyle A}{B} }

\newcommand{\AB}{{A,B}}
\newcommand{\BA}{{B,A} }
\newcommand{\Ord}{{\cal O}}
\newcommand{\ms}{\mw{\sigma}}
\newcommand{\mdts}{\mw{\dts}}

\newcommand{\resto}[1]{\left.\right|_#1}
\newcommand{\sA}{\sigma_A}
\newcommand{\sB}{\sigma_B}
\newcommand{\dtsb}{\dts_B}
\newcommand{\teta}{\Theta_A}

\newcommand{\lap}{\raisebox{-0.5ex}{$\stackrel{\scriptstyle <}{\sim}$}}
\newcommand{\gap}{\raisebox{-0.5ex}{$\stackrel{\scriptstyle >}{\sim}$}}
\newcommand{\csp}{{c_{\rm s}}}
\let\tet=\vartheta
\newcommand{\vaq}{\bar{v}_A}
\newcommand{\oaq}{\bar{\omega}_A}

\newcommand{\dreiHe}{\raisebox{1ex}{\scriptsize 3}He}

\newcommand{\pdf}[1]{\partial_{#1}}
\newcommand{\pdfn}[2]{\partial_{#1}^{#2}}

\expandafter\ifx\csname mathrm\endcsname\relax
  \def\mathrm#1{{\rm #1}}\fi

\def\ITPLogo{
\setlength{\unitlength}{10pt}%
\begin{picture}(10,11)
\put( 1,6){\line( 0,1){ 3}}
\put( 2,6){\line( 0,1){ 3}}
\put( 3,0){\line( 0,1){ 8}}
\put( 3,9){\line( 0,1){ 1}}
\put( 2,10){\line( 0,1){ 1}}
\put( 4,1){\line( 0,1){ 1}}
\put( 4,3){\line( 0,1){ 1}}
\put( 5,1){\line( 0,1){ 1}}
\put( 5,3){\line( 0,1){ 1}}
\put( 4,5){\line( 0,1){ 5}}
\put( 5,5){\line( 0,1){ 5}}
\put( 6,0){\line( 0,1){ 8}}
\put( 6,9){\line( 0,1){ 1}}
\put( 7,6){\line( 0,1){ 3}}
\put( 8,8){\line( 0,1){ 1}}
\put( 9,8){\line( 0,1){ 1}}
\put( 10,8){\line( 0,1){ 1}}
\put( 8,6){\line( 0,1){ 1}}
\put( 7,10){\line( 0,1){ 1}}
\put( 1,10){\line( 1,0){ 1}}
\put( 0,6){\line( 1,0){ 1}}
\put( 2,6){\line( 1,0){ 1}}
\put( 0,9){\line( 1,0){ 1}}
\put( 2,9){\line( 1,0){ 1}}
\put( 3,10){\line( 1,0){ 1}}
\put( 2,11){\line( 1,0){ 5}}
\put( 4,5){\line( 1,0){ 1}}
\put( 5,10){\line( 1,0){ 1}}
\put( 4,4){\line( 1,0){ 1}}
\put( 4,3){\line( 1,0){ 1}}
\put( 4,2){\line( 1,0){ 1}}
\put( 6,9){\line( 1,0){ 1}}
\put( 6,6){\line( 1,0){ 1}}
\put( 7,10){\line( 1,0){ 2}}
\put( 8,9){\line( 1,0){ 1}}
\put( 8,8){\line( 1,0){ 1}}
\put( 8,7){\line( 1,0){ 1}}
\put( 8,6){\line( 1,0){ 1}}
\put( 3,8){\line( 1,1){ 1}}
\put( 3,0){\line( 1,1){ 1}}
\put( 0,9){\line( 1,1){ 1}}
\put( 5,4){\line( 1,1){ 1}}
\put( 6,9){\line( 1,1){ 1}}
\put( 9,7){\line( 1,1){ 1}}
\put( 6,3){\line( 1,1){ 3}}
\put( 2,10){\line( 1,-1){ 1}}
\put( 3,5){\line( 1,-1){ 1}}
\put( 5,9){\line( 1,-1){ 1}}
\put( 9,10){\line( 1,-1){ 1}}
\put( 5,1){\line( 1,-1){ 1}}
\put( 0,6){\line( 1,-1){ 3}}
\end{picture}
}
\begin{document}

\title{
       Bubble Growth in Superfluid \dreiHe:\\
       The Dynamics of the Curved A-B Interface
      }
\author{
        Jens Johannesson\thanks{e-mail:
         {\tt johan@kastor.itp.uni-hannover.de}}
        and Mario Liu\\
        \\
	\ITPLogo
        \\
        {\normalsize Institut f\"ur Theoretische Physik,
                     Universit\"at Hannover}\\
        {\normalsize D-3000 Hannover 1, F.~R.~Germany}
       }
\date{June 25, 1993}
\maketitle

PACS-numbers: 67.57.-z, 67.55.Fa \hfill ITP-UH 11/93

\begin{abstract}
We study the hydrodynamics of the A-B interface with finite curvature.
The interface tension is shown to enhance both the transition velocity
and the amplitudes of second sound. In addition, the magnetic signals
emitted by the growing bubble are calculated, and the interaction
between many growing bubbles is considered.
\end{abstract}

\newpage
\section{Introduction}\label{sec:int}
The interface between the A- and B-phase of \dreiHe\ is the only known
boundary between two superfluid systems. The interface motion that
accompanies the first order phase transition from the undercooled
A- into the B-phase is subject to extended experimental
\mbox{[1 -- 4]}
and theoretical
\mbox{[5 -- 9]}
investigation.

Within the hydrodynamic regime, the structure of the interface
dynamics can be rigorously determined by connecting the hydrodynamic
modes on both sides of the interface with the appropriate connecting
conditions (CoCos). This has already been done
\cite{GLab,panzspinshort,KoL} for the planar interface.  In the
present work, we generalize the hydrodynamic description to interfaces
of finite curvature, taking as an example spherical B-phase bubbles
growing in a bath of undercooled A-phase.

To gain a qualitative understanding, consider first planar growth. For
small growth velocities, the released latent heat serves as the source
for two second sound step functions, emitted in opposite directions
\cite{GLab}. The steps move with $c_{2A}$ and $-c_{2B}$ if the A-phase
is located on the right and $c_2$ denotes the second sound velocity.
If the interface velocity $\dot{u}$ becomes larger than $c_2$, both
steps are left behind in the B-phase. The same is true for spin waves
if an external magnetic field produces excess magnetization at the
interface \cite{panzspinshort}. The reference velocity is now of
course the spin wave velocity $\csp$, rather than $c_2$.

Qualitatively, bubble growth is not much different, except that the
surface tension delivers an extra push towards the energetically more
favorable B-phase: Neglecting the anisotropy, radial step functions of
second sound are emitted out- and inward for $\dot{u}\ll c_2$. The
inward moving step functions quickly equalize the temperature
distribution within the B-phase bubble. For $\dot{u}\gg c_2$, both
radial waves of second sound remain inside the bubble while the
temperature remains homogeneous in the A-phase. The spin wave and the
magnetization behave accordingly. (Though spin waves are emitted only
for higher interface velocities $\dot{u}\gap\,1 m/s\,\gg\,c_2$
\cite{panzspinshort}) The case $\dot{u}\gg \csp$ may seem academic but
one really cannot discount the possibility that it was indeed already
observed \cite{slowgrowth,panzspinshort}. Then, even both spin waves
remain inside the bubble where neither the temperature nor the
magnetization is homogeneous. Causality sees to it that the A-phase is
completely unperturbed.

The complication of bubble growth is in its time dependence.  While
planar growth keeps both $\dot{u}$ and the amplitudes of the emitted
waves constant, neither are in bubble growth. The growth velocity is
not constant because the pushing surface tension, proportional to the
inverse of the bubble radius $R(t)$, is not. The amplitudes are not
constant, because first of all, they vary as $1/r$ when moving away
from the interface; second, the amplitudes in addition are functions
of $\dot{R}(t)$. (In fact, there is a retardation effect, the
amplitudes are functions of $\dot{R}(t-t_f)$, where $t_f$ is the time
of flight $t_f=(r-R)/c_2$ or $(r-R)/\csp$.)

The anisotropy of $c_2$ and $\csp$ will distort the radial wave into
an elliptical shape, with the principal axes being parallel and
perpendicular to $\hat{\ell}$, the orbital preferred direction. (We
presume $\hat{\ell}$ to be uniform.) In addition, the growth and
Kapitza coefficients are also $\hat{\ell}$-dependent, and this
directly distorts the bubble. (Given enough time, of course, the
surface tension will drive the bubble into the shape that minimizes
the surface energy; generally, it is somewhere inbetween, reflecting
some stationary balance between the surface tension and the uneven
growth.) Assuming the same eccentricity for both ellipses, we can
rescale to obtain isotropic growth and radial waves to perform the
calculations.  Afterwards, the distortion into elliptical waves is
achieved by taking the correct value of the anisotropic velocity for
the two principle axes.

If there are many bubbles present, their growth can be considered
completely independent for $\dot{R}>c_2$, because the magnetic signals
barely feed back to the growth \cite{panzspinshort}, while the second
sound waves, which do feed back, are confined to the interior of the
bubbles. For $\dot{R}< c_2$ however, the many emitted second sound
waves heat up the surrounding A-phase. Since the latent heat is not
completely evacuated, the growth velocity slows down accordingly.

A B-phase bubble is thermodynamically stable only if its radius
exceeds the critical value $R_c=\alpha/(\rho \, \Delta \mu$)
\cite{YLabdyn,Rem1}, where $\alpha$ denotes the interface tension and
$\Delta \mu= \mu_B - \mu_A$ is the difference in the chemical
potential across the interface (for every quantity $X$, we define:
$\mw{X}=(X_A+X_B)/2$ and $\Delta X=X_B-X_A$). Since $\Delta \mu$
vanishes at the coexistence point, $R_c$ diverges if $T$ approaches
$T_{AB}$. At lower temperatures, \mbox{$R_c < l_{QP}$}, where $l_{QP}$
is the mean free path of the quasi particles, the hydrodynamic
description becomes valid only after the bubble has nucleated and
reached an appropriate size by ballistic processes. Fig.\ 1 shows the
respective length scales as a function of temperature. As pointed out
by Leggett and Yip \cite{YLabdyn} the nucleation process itself is
likely to be of exotic nature.

The CoCos \cite{GLab,panzspinshort,KoL}, we shall employ fall into
three categories, those for the conserved quantities, those for the
symmetry variables and the parameterization of the surface entropy
production. For the locally conserved quantities, the CoCos state the
continuity of the respective currents, expressing the fact that
conserved quantities cannot accumulate within, or be depleted from, a
region of microscopic width. This is not true, however, if the
momentum current traverses a curved interface, where the radial bulk
momentum current differs by the interfacial momentum current
\cite{Rem1} $\alpha/R$. (Note that $\alpha<0$: In equilibrium the
pressure is larger on the A-phase side.) With prime denoting the rest
frame of the interface, the CoCos are:
\begin{eqnarray}
\Delta g'&=&0, \label{rbmass}\\
\Delta (p + \Pi')&=&\frac{\alpha}{R},  \label{rbmomentum}\\
\Delta Q' &=& 0, \label{rbq} \\
\Delta {\bf j}^{\rm spin}&=&0. \label{rbspin}
\end{eqnarray}
The notations: mass current $g'$, pressure $p$, nonlinear part of the
momentum current $\Pi'$, energy current $Q'$ and spin current
${\bf j}^{\rm spin}$.

The behavior of the symmetry variables in the interface region is
related to the elasticity of the order parameter within the interface.
Since the elastic coefficients are of comparable order of magnitude in
the interface as in the bulk liquid, we have coherence of the symmetry
variables across the interface. More specifically, we have for the
phase angle $\phi$ and the rotation angle $\vec{\Theta}$ in spin space
\begin{eqnarray}
\Delta\,\phi&=&0 ,\label{rbphi}\\
\Delta\,\vec{\Theta}&=&0.\label{rbtet}
\end{eqnarray}
It is by means of Eq.\ (\ref{rbphi}) that the interface tension acts
in accelerating the growth process. Since (\ref{rbphi}) implies
$\Delta \mu = 0$, the interface tension couples to the dynamics through
the pressure dependency of the chemical potential and CoCo
(\ref{rbmomentum}) is implicitly incorporated. When the growth starts
at $R=R_c$ we find the initial interface velocity to be twice as fast
as in the planar case. This effect should be accessible to
experimental probe.

The irreversibility of the phase transition is accounted for by the
growth coefficient $K$ for $\dot{u}\gg c_2$, and by the Kapitza
resistance $\kappa$ for $\dot{u}\ll c_2$, respectively.  Both $K$ and
$\kappa$ are effective, integrated quantities and functions of various
surface Onsager coefficients, thermodynamic susceptibilities and bulk
transport coefficients \cite{KoL}. The rate of interface entropy
production $R_S=-\mw{T}\,\Delta\,f'$ is then respectively given as
\begin{equation} \label{rsentro}
R_S = \frac{{g'}^2}{K} \qquad \mbox{or} \qquad
                     R_S = \frac{{\mw{f'}}^2}{\kappa},
\end{equation}
where $f'$ is the entropy current. So except Eq.\ (\ref{rbmomentum}), all
CoCos retain their form of planar growth.

The solution of these CoCos in Sec.\ \ref{sec:dyn}
shows the following result: The growth velocity decreases with $1/R$
towards the limiting value $\Rpinf$ (of the planar growth); while the
amplitudes of second sound generated during the growth show some
intricate time dependency.

In Sec.\ \ref{sec:spin}, we discuss the magnetic signals emitted
by the growing bubble. These are more accessible to experimental
probe.

The general case of $N$ interacting bubbles $V$ is discussed in Sec.\
\ref{sec:nbub}.

\section{Spin waves}\label{sec:spin}
We consider the magnetic signals emitted during bubble growth for
$\dot{R}\ll\csp$. As discussed in Ref.\ \cite{panzspinshort}, the
magnetization is for
$\dot{R}\lap\,(H/\mathrm{kOe})^{-1}\,{\mathrm{m}/\mathrm{s}}$
essentially given by its equilibrium values $\chi_{A,B}\,H$. The small
deviations from the equilibrium values do not change much, going from
planar to bubble growth. For higher velocities, $\dot{R}\gap
1\,\mathrm{m}/\mathrm{s}$, the dipole-interaction is negligible, the
magnetic signals turn into spin waves and become delocalized. Here,
the circumstances of bubble growth are more complicated and are what
we shall consider.

\subsection{Interface dynamics}\label{subsec:intdyn}
The feed-back of the spin-dynamics to the growth velocity $\dot{R}$ is
negligible if $\dot{R}$ is not too close to $\csp_A$
\cite{panzspinshort}. Hence, we may take $\dot{R}$, as calculated in
Sec.\ \ref{sec:dyn} without the inclusion of the magnetic fields, as
an input:
\begin{equation} \label{Rpvt}
\dot{R}=\Rpinf\,\left(1+\frac{R_c}{R}\right),
\end{equation}
where $\Rpinf$ denotes the limiting value of the interface velocity
for large $R$, i.e.\ $\Rpinf=\dot{u}$ of planar growth.  The explicit
values for $\Rpinf$ are given in Eqs.\ (\ref{rpinfslow}) and
(\ref{rpinffast}) for weak and strong undercooling, respectively. The
solution of the differential equation (\ref{Rpvt}) subject to the
initial condition $R(t=0)=R_0$ then reads
\begin{equation} \label{tauvx}
t\,\Rpinf = R-R_0+R_c\,\ln\left(\frac{R_0+R_c}{R+R_c}\right).
\end{equation}
The inverse of (\ref{tauvx}) can also be written in a closed form, see
App.\ \ref{app:xvtau}.

\subsection{Spin-Hydrodynamics}
We shall work with the longitudinal model \cite{panzspinshort}, which
contains only two dynamical variables: the magnetization and the
dipole angle $\Theta$ (where $\Theta_A$ is that between $\hat{d}$ and
$\hat{\ell}$, and $\Theta_B$ is the angle around $\hat{n}$).

The initial scenario contains a B-phase bubble of radius $R_0$,
exposed to a homogeneous magnetic field $\vec{H}$. Within the bubble,
we assume a uniform $\hat{n}$-texture, parallel to $\vec{H}$; in the
surrounding A-phase we have $\hat{d}\perp\vec{H}$,
$\hat{\ell}\parallel\hat{d}$. Although the uniformity of the
$\hat{n}$-texture contradicts the equilibrium A-B interface
\cite{thuneeqbc}, we shall neglect this complication as in Ref.\
\cite{panzspinshort}.

The coordinates are chosen such that $\vec{H}$ points in the
z-direction, while the x-axis coincides with $\hat{\ell}$.  The
spin-hydrodynamics in the A-phase is then given by the wave equation
\begin{equation} \label{sebwgl}
\ddot{\Theta}_A - \csp_\|^2 \partial_x^2\, \teta
  - \csp_\perp^2(\partial_y^2 + \partial_z^2)\,\teta =0,
\end{equation}
where $\csp_\|$ and $\csp_\perp$ denote the spin wave velocities
parallel and perpendicular to the preferred direction $\hat{\ell}$.
The consequences of neglecting spin diffusion will be discussed at the
end of this section.

Eq.\ \ref{sebwgl} shows that the outgoing wave fronts will be of
ellipsoidal shape with a rotational symmetry around $\hat{\ell}$.
Assuming a growing bubble with the same eccentricity, as discussed in
the introduction, we may distort the coordinates to render both
ellipsoids (interface and wave front) spherical:
\begin{equation} \label{sebdyn}
\ddot{\Theta}_A- \csp_A^2 \Delta_r \teta =0, \quad \omega_A=\dot{\Theta}_A,
\quad v_A=\pdf{r}{\Theta_A}.
\end{equation}
($v_A$ is the spin-superfluid velocity.)

The B-phase spindynamics need not be considered explicitly. After the
initial time lag of $R_0/\csp_B$, the imploding wave front meets at
the center. There, the excess magnetization (always parallel to the
external field $\vec{H}$) from opposite directions add up, while the
spin superfluid velocities $\nabla\,\Theta_B$ (all having the same
sign along the radial direction) from opposite directions cancel. So
homogeneous magnetization and vanishing $\nabla\,\Theta_B$ is a good
approximation.

\subsection{Connecting Conditions}
With $\dot{R}\ll\csp_A$, we can safely neglect terms of order
$\dot{R}\,v_A$ and $\dot{R}\,\delta S$; the CoCos
(\ref{rbtet}) and (\ref{rbspin}) therefore take the simplified form
\begin{eqnarray}
0&=&\Delta \omega, \label{sebbc1} \\
0&=&\gamma H \dot{R} \Delta \chi - \chi_A \csp_A^2 v_A, \label{sebbc2}
\end{eqnarray}
with gyromagnetic ratio $\gamma$ and susceptibility $\chi$. Working
with the longitudinal model, only one of the three relations
(\ref{rbtet}) is needed here.

\subsection{Solution of the Connecting Conditions}
The procedure to determine the space and time
dependency of the radial waves from their CoCos is
given below:

In the first step, we calculate the values  of the general solution
\begin{equation} \label{gensolteta}
\Theta_A= \frac{f(r-\csp_A\,t)}{r}
\end{equation}
at the interface (subscript $I$) from the
CoCos (\ref{sebbc1}, \ref{sebbc2}).

Substituting the time dependent bubble radius $R(t)$ for $r$, we find
\begin{eqnarray}
\omega_A\resto{I}&=& \left( \pdf{t} \Theta_A(r,t)\right)\resto{{r=R(t)}}
= -\csp_A \left.\frac{f'(r-\csp_A\,t)}{r}\right|_{r=R(t)},\label{omegainter}\\
v_A\resto{I}&=& \left( \pdf{r}\Theta_A(r,t)\right)\resto{{r=R(t)}}
= \left.\left(\frac{f'(r-\csp_A\,t)}{r} - \frac{f(r-\csp_A\,t)}{r^2}\right)
  \right|_{r=R(t)}. \label{vinter}
\end{eqnarray}
where $f'$ denotes the derivative of $f$ with respect to its argument.
Now, the amplitudes at the interface are functions of time only.  For
sake of clarity, we introduce an auxiliary function by
\begin{equation} \label{defF}
F(t)=f\left(R(t)- \csp_A\,t\right)\equiv f\left(\bar{R}(t)\right),
\end{equation}
implying
\begin{equation} \label{defFprime}
f'= \frac{\pdf{t} F(t)}{\dot{R} - \csp_A}.
\end{equation}
The CoCos (\ref{sebbc1}, \ref{sebbc2}) thereby turn into differential
equations for $F(t)$ which can be solved.

Having obtained $F(t)$ with the appropriate initial condition
$F(t=0)=0$, the next step is to reconstruct the $(r,t)$-dependency of
$f$, where we need the inverse function of $\bar{R}(t)$:
\begin{equation} \label{Fvt1}
F\left(t(\bar{R})\right)=f(\bar{R}).
\end{equation}
Given $f(\bar{R})$, we can substitute $\bar{R}=r-\csp_A\,t$ and obtain
$\omega_A$ and $v_A$ for all $r$, $t$ via Eq.\ (\ref{sebdyn}).

For technical reasons, however, we modify this procedure and eliminate
the time $t$ with the inverse of $R(t)$, instead of eliminating $r$ by
$R(t)$, as described above. The reader is referred to App.\
\ref{app:spincalc} for the details of the calculation and the
result.

For the qualitative features cf.\ Figs.\ 2 and 3. The spatial
variation of the outgoing radial wave is illustrated for different
times. Each curve displays two effects. First the $1/r$-decrease as
the wave moves out, and second the amplitude at $r,t$ is a function of
$\dot{R}$ at $t=(r-R)/c$, a retardation effect.  The line at the upper
surface of the box marks the location of the bubble surface, relative
to the starting value. Below, one can track the decrease of the
amplitudes at the interface.  As will be shown in Sec.\ \ref{sec:dyn},
these plots also visualize the second sound amplitudes, emitted by a
slowly growing bubble.

At the head of the shock wave, i.e. where $r=R_0+\csp_A\,t$, the
expressions for the amplitudes simplify to become
\begin{equation} \label{sebhead}
v_A(t)=\frac{\gamma H}{\csp_A}\frac{\Delta \chi}{\chi}\frac{\Rpinf}{\csp_A}
 \frac{R_0/R_c(1+R_0/R_c)}{R_0+\csp_A\,t},\quad
\omega_A(t)=-\csp_A\,v_A(t).
\end{equation}

The qualitative effect of the spin diffusion, neglected here, is the
rounding of the magnetization step with $\omega\,t$ at the wave front.
So Eq.\ (\ref{sebhead}) is valid for the total change in $v_A$ and
$\omega_A$.

\section{Growth Dynamics of spherical bubbles}\label{sec:dyn}
The interface dynamics of superfluid \dreiHe\ is fundamentally
different from other systems \cite{GLab}. If $\dot{u}\ll c_2$, the
evacuation of latent heat is very efficiently accomplished via second
sound, the difference between hypercooling and normal supercooling is
inessential, and $\dot{u}$ is determined by the Kapitza resistance of
the interface. For $\dot{u}\gg c_2$, the system is always hypercooled,
no heat evacuation needs to take place and not much does. $\dot{u}$ is
determined by the growth coefficient.

In bubble growth, two similar scenarios prevail: For weak undercooling
of the A-phase, $\dot{R}\ll c_2$, an outgoing radial wave is emitted
into the A-phase, while an imploding wave is emitted into the bubble.
The latter will interfere at the center of the bubble, as in the case
of spin waves, yielding vanishing counterflow and an averaged
temperature, depending only on the current bubble radius.

For strong undercooling, assuring $\Rp \gg c_2$ at all time, no sound
is emitted into the A-phase, while an exploding and an imploding wave
superpose inside the bubble, yielding a time dependent temperature
distribution.

\subsection{Bulk Hydrodynamics}
We start from the linearized equations of bulk hydrodynamics (correct
for both limits \cite{GLab} $\dot{R}\gg c_2$, $\dot{R}\ll c_2$ but not
in-between, $\dot{R}\approx c_2$ \cite{jjmlnonlin}), in a form
appropriate for the spherical symmetry of the problem. As in planar
growth, first sound is too fast to need explicit consideration. Its
only role is to maintain uniform pressure on both sides of the
interface, while keeping the discontinuity $\Delta\,p=\alpha/R$ across
it. This leaves two equations: Conservation of entropy and the
equation for the counterflow. The counterflow,
$\vec{w}=w(r)\,\hat{e}_r$ is curl-free and can be written as
$w(r)=\pdf{r}\phi$. The two hydrodynamic equations then become
\begin{eqnarray}
\frac{\rho \sigma}{\rn}\,\dot{\delta T} + c_2^2\,\Delta_r \phi
     &=&0, \label{spherentro}\\
\pdf{r}\left(\dot{\phi} + \frac{\rho \sigma}{\rn}\,\delta T\right)
     &=&0. \label{sphersymvar}
\end{eqnarray}
Eq.\ (\ref{sphersymvar}) seems to allow $\dot{\phi}$ and
$-\rho\sigma/\rn\,\delta T$ differing by an explicit function of time.
This is ruled out by the wave equation that can be obtained from the
combination of (\ref{spherentro}) and (\ref{sphersymvar}), which must
not have any time dependent source.

In summary, we end up with
\begin{equation} \label{secsndweq}
w=\pdf{r}\,\phi(r,t),\quad \delta T=-\frac{\rn}{\rho\sigma}
  \dot{\phi}(r,t),\quad \ddot{\phi}-c_2^2\Delta_r\,\phi=0,
\end{equation}
in close analogy to (\ref{sebdyn}).

\subsection{Weak undercooling}
We will consider two different stages: The first stage is
characterized by the existence of an incoming and an outgoing second
sound wave and lasts only for a time span of order $R_0/c_2$, where
$R_0\equiv R(t=0)$. The second stage starts after the interference has
established homogeneity inside the bubble, making the temperature
solely dependent of the bubble radius. In the calculation of the
second stage we shall neglect the existence of the first. This only
changes the initial condition and hence the exact shape of the radial
wave front but not its magnitude. For very weak undercooling, $R_c$ is
large compared to the quasiparticles mean free path, so that for $t=0$
$R=R_c$ and $T=T_i$ are the valid initial conditions. Otherwise, we
assume a nonvanishing initial temperature discontinuity
$\Delta\,T_i=T_{iB}-T_{iA}$ and $R=R_0\gg R_c$ at $t=0$.

We expand the CoCos (\ref{rbq},\ref{rbphi},\ref{rsentro}), to first
order in $\Rp$, $\delta T_{\AB}$, $w_{\AB}$ around the initial
temperature $T_i$:
\begin{eqnarray}
0&=& -\rho \Rp \delsig + \rs \ms (w_B -w_A), \label{sbc1} \\
0&=& \delmu (1+\frac{R_c}{R}) - \ms (\dtb + \DTi- \dta), \label{sbc2} \\
0&=& -\rho \Rp \ms - \kappa (\dtb + \DTi - \dta), \label{sbc3}
\end{eqnarray}
all quantities refer to the laboratory frame. Terms of Order
$(\delsig/\ms)^2$ as well as $\pdf{p}(\rs/\rho)$, $\pdf{T}(\rs/\rho)$
and $\Delta \rho/\rho$ are omitted.

{}From (\ref{sbc2}) and (\ref{sbc3}) we readily obtain the interface
motion as quoted in Eq.\ (\ref{Rpvt}). The limiting value of the
interface velocity is
\begin{equation} \label{rpinfslow}
\Rpinf=-\frac{\kappa\,\delmu}{\ms^2\,\rho}.
\end{equation}

We consider the
sound amplitudes at the interface for the first stage. Starting with
\[
\phi(r,t)=\frac{f(r-c_2\,t)}{r} + \frac{g(r-c_2\,t)}{r},
\]
we apply the method described in Sec.\ \ref{sec:spin}
and calculate the sound amplitudes to lowest order in
$(R-R_c)/R_c$:
\begin{eqnarray}
\delta T_{\staAB} &=& \mp \frac{\delmu}{\ms}
 + \frac{1}{2}\DTi
 - \frac{\Rpinf}{c_2}\frac{\delsig}{\mdts} \\ \nonumber
&& + \frac{1}{2}\left\{\frac{\delmu}{\ms}\left(\frac{c_2}{\Rpinf}\pm 1\right)
 -\frac{1}{2}\DTi\frac{c_2}{\Rpinf}
 -\frac{\Rpinf}{c_2}\frac{\delsig}{\ms}\right\}\frac{R-R_c}{R_c}
 +\Ord\left(\left(\frac{R-R_c}{R_c}\right)^2\right), \label{sdT} \\
w_{\staAB}&=& \frac{\rho\ms}{\rn c_2}\left[-\frac{\delmu}{\ms}
 +\frac{1}{2}\DTi
 \mp\frac{\Rpinf}{c_2}\frac{\delsig}{\mdts} \right. \\ \nonumber
&& \left. + \frac{1}{2}\left\{\frac{\delmu}{\ms}\left(\pm2
 \frac{c_2}{\Rpinf}+3\right) - \DTi\left(\pm\frac{c_2}{\Rpinf} + 1\right)
 -\frac{\delsig}{\mdts}\left(1\pm3\frac{\Rpinf}{c_2}\right)\right\}
 \frac{R-R_c}{R_c} \right.\\ \nonumber
&&\left. + \Ord\left(\left(\frac{R-R_c}{R_c}\right)^2\right)
 \right]. \label{sw}
\end{eqnarray}
Setting $\DTi$ and $R-R_c$ to zero, these expressions can be compared
to the results of planar growth \cite{GLab}: As it turns out, the
surface tension doubles the initially emitted sound amplitudes. Note
that the terms of first order in $R/R_c-1$ are dominated by the
contribution proportional to $c_2/\Rpinf$.

For the second stage, the situation is quite similar to that of
spinwaves, examined in the last section. Since (\ref{sbc2}) and
(\ref{sbc3}) have already been used once to determine the bubble
radius as a function of time, we shall now only consider the CoCos
(\ref{sbc1}) and (\ref{sbc2}). With $w_B$ vanishing, (\ref{sbc1})
simplifies to
\begin{equation} \label{aeq1}
0=\rho\,\delsig\,\dot{R} + \rs\ms\,w_A,
\end{equation}
which is analog to (\ref{sebbc2}) when $w_A$ is identified with
$-v_A$. Combining (\ref{aeq1}) with (\ref{sbc2}) yields
\begin{equation} \label{aeq2}
w_A= \frac{\kappa\,\delsig}{\ms^2\,\rs}\left(\dtb - \dta + \DTi \right).
\end{equation}
By identification of $\dta$ with $\omega_A$, (\ref{aeq2}) is analogous
to (\ref{sebbc1}). (These identifications are justified by inspection
of the bulk hydrodynamics: (\ref{secsndweq}) goes over into
(\ref{sebdyn}) by the substitution of $-\Theta_A$ for $\phi$; but this
does not alter the solution of the wave equation.)

We can therefore take over the results of Sec.\ \ref{sec:spin} by
setting
\begin{eqnarray}
w_A&=&-v_A, \label{waaussen}\\
\dta&=&\frac{\rn}{\rho\ms}\frac{c_{2A}}{\csp_A}\omega_A, \label{dTaaussen}
\end{eqnarray}
when we in addition observe the respective redefinition of $A$ and
$c_A$,
\begin{equation} \label{redefAcA}
A=-\frac{\delsig}{\ms}\frac{\rs}{\rho}\Rpinf\,R_c, \quad
c_A= \frac{c_{2A}}{\Rpinf},
\end{equation}
introduced in App.\ \ref{app:spincalc}. As a consequence of this
analogy, Figs.\ 2 and 3 also serve as illustrations of counterflow and
temperature amplitudes.

\subsection{Strong undercooling}
At the strong degree of undercooling we are considering here, the
hydrodynamic description only holds after the bubble has grown to an
appropriate multiple of the critical size. We therefore impose the
initial condition $R(t=0)=x_0\,R_c$ with $x_0\gg 1$.

Inside the bubble, an incoming and an outgoing second sound wave
superpose. After the incoming wave front has been ``reflected'' at the
center of the bubble, there are two spherical discontinuities of
temperature and counterflow, expanding outward and laging behind the
third discontinuity, at the bubble surface.

Neglecting terms of the order $w_B/\Rp$, an expansion of
(\ref{rbq},\ref{rbphi},\ref{rsentro}), again in $R_c/R-1$, yields for
the discontinuities at the bubble surface
\begin{eqnarray}
0&=& T_i \delsig + \delmu \left(1+\frac{R_c}{R}\right)
                 + T_i\, \dtsb\, \left(\dtb+\DTi\right), \label{fbc2} \\
0&=& \delmu \left(1+\frac{R_c}{R}\right) -
            \sB \left(\dtb + \DTi\right)
            +\frac{\rn}{\rho}\Rp\, w_B, \label{fbc1} \\
\frac{\rho\Rp}{K} &=& T_i \delsig
    + \left(T_i \dtsb + \frac{\delsig}{2}\right)
    \left(\dtb+\DTi\right). \label{fbc3}
\end{eqnarray}
{}From (\ref{fbc2}) we obtain the temperature
\begin{equation} \label{dtf}
\dtb=-\DTi-\frac{\delmu}{T_i\dtsb}\left(\frac{1+R/R_c}{R/R_c}\right)
     -\frac{\delsig}{\dtsb},
\end{equation}
which, together with (\ref{fbc3}), again leads to (\ref{Rpvt}). Here
the limiting value of the interface velocity is
\begin{equation} \label{rpinffast}
\Rpinf=-\frac{K\delmu}{\rho}\left(1+\frac{1}{2}\frac{\delsig}{T_i\dtsb}\right).
\end{equation}

By calculation of the total counterflow amplitude
\begin{equation} \label{wbf}
w_B=-\frac{\rho}{\rn}\left\{\frac{\delmu}{\Rpinf}
    \left(1+\frac{\sB}{T_i\dtsb}\right)
    + \frac{\sB\delsig}{\dtsb}\frac{1}{\Rp}\right\},
\end{equation}
our initial neglect of $\Ord(w_B/\Rp)$ proves to be appropriate.

\section{Interacting Bubbles}\label{sec:nbub}
As mentioned in the introduction, we expect the growth velocity of the
bubble to slow down, when its surrounding heats up from second sound
radiation sent out by other bubbles. Only the case of weak
undercooling ($\Rp < c_2$) is worth studying, since for strong
undercooling ($\Rp > c_2$) all the latent heat remains inside the
bubble.

Let us for simplicity consider a volume $V$ of A-phase, containing $N$
B-phase bubbles, nucleated simultaneously and distributed uniformly
over $V$. Each individual bubble grows as described in the preceding
section until it is reached by the second sound pulses of its
neighbors. {}From now on, the accumulative heating of the A-phase
increasingly hinders the evacuation of latent heat and slows down the
growth velocity.

We will restrict our calculations to the regime where the temperatures
of both the A- and B-phase can be regarded as being uniform. The
Volume $V$ is thus divided up into two subsystems: $N$ bubbles of
total volume $V_o=N 4 \pi /3 R^3$ and the remaining A-phase of volume
$V-V_o$. Since the gradients vanish, the local hydrodynamic
description of the previous sections reduce to thermodynamic
considerations: For small deviations from equilibrium, energy
conservation accounts for equality of the entropy currents,
\begin{equation} \label{nbc1}
\dot{S}_B=-\dot{S}_A,
\end{equation}
where $S_B=V_o\,\rho\,\sB$ and $S_A=(V-V_o)\,\rho\,\sA$ denote the total
entropies of the respective phase. Because of the identity
\begin{equation} \label{surfcurrent}
d_t\,S_B=\int\limits_{V_o} \!\!dV\,\dot{\sigma}_B
  + \int\limits_{\partial V_o} \!\!\! ds \, \dot{R}\sigma_B
= -\int\limits_{\partial V_o} \!\!\! ds \,{f'}_B,
\end{equation}
we identify Eq.\ (\ref{nbc1}) as the continuity condition of the local
entropy current, integrated over the area of the interface. In the
same manner, we obtain the Onsager relation, starting from
${f'}_B=\kappa\,\Delta T$ as
\begin{equation} \label{nbc2}
-\frac{\dot{S}_B}{A_I}=\kappa\,(\dtb - \dta).
\end{equation}
Here, $A_I$ denotes the area of the interface so that the total rate
of interface entropy production is an extensive function with respect
to the interface area, as it must be. Being of intensive nature, the
continuity condition for the phase angle
\begin{equation} \label{nbc3}
0=\delmu\left(1+\frac{R_c}{R}\right) - \sB\,\dtb + \sA\,\dta
\end{equation}
does not change its appearance.

Because of the homogeneity of the subsystems, all thermodynamic
quantities are explicit functions of $R$. We therefore need to specify
only a single initial condition:
\begin{equation} \label{nbini}
\dtb(R_i)-\dta(R_i)=\Delta\,T_i,
\end{equation}
where $R_i$ denotes the initial bubble radius and $\Delta\,T_i$ is the
temperature difference that was built up across the interface during
the switch-on process. {}From Sec.\ \ref{sec:dyn}, we expect
$\Delta\,T_i$ to be of the order $\delmu/\ms$. Since (\ref{nbc3}) was
already valid before homogeneity had been established, the initial
temperature amplitudes are found to be
\begin{equation} \label{Tempini}
\dta(R_i)=\frac{\delmu-\sB\,\Delta T_i}{\delsig},\quad
\dtb(R_i)=\frac{\delmu-\sA\,\Delta T_i}{\delsig}.
\end{equation}
Here and below, we neglect the effect of interface tension, i.e.\ we
assume $R_c \ll R_i$.

{}From (\ref{nbc1}) and the derivative of (\ref{nbc3}) with respect to
$R$ we obtain $\pdf{R}\dta$ and $\pdf{R}\dtb$ which may be integrated
to give the temperature amplitudes as functions of the bubble radius:
\begin{equation} \label{TempvR}
\delta T_{\AB}(R) = \delta T_{\AB}\,(R_i) - \frac{\sigma_{\BA}\,
                      \delsig}{a}\,\ln\left(\frac{4\pi/3 N R^3\,a
                      + V\,b}{4\pi/3 N R_i^3\,a + V\,b}\right),
\end{equation}
with $a=\sA\,\dts_B-\sB\,\dts_A$, $b=\sB\,\dts_A$.

The growth velocity follows from (\ref{nbc2}):
\begin{equation} \label{nbRp}
\dot{R}(R)=-\frac{\kappa}{\rho}\Delta T_i
  \frac{\dtb(R)-\dta(R)}{\sB+\dts_B\,R\,\pdf{R}\dtb/3}.
\end{equation}

Obviously, the system approaches an equilibrium state with equal
temperatures on both sides of the interface and vanishing $\dot{R}$.
Starting with
\[
\dot{R}(R_i)\approx - \frac{\kappa}{\rho}\frac{\Delta T_i}{\ms} \left(
       1-\frac{\delsig}{\ms}\left\{\frac{1}{2}-
       \frac{4\pi/3 N R_i^3}{V}\right\}\right),
\]
the interface velocity decreases and eventually comes to rest at the
final bubble radius $R_f$. The B-phase bubbles then occupy the volume
\begin{equation} \label{finVol}
\frac{4\pi}{3} N R_f^3 = \frac{1}{a}\left\{
\left(\frac{4\pi}{3}N R_i^3\,a + V\,b\right) \exp\left(-
\frac{a\,\Delta T_i}{\delsig^2}\right) - V\,b
\right\}
\end{equation}
and the final temperature amplitudes amount to
\begin{equation}
\dta(R_f)=\dtb(R_f)=\frac{\delmu}{\delsig}.
\end{equation}
\appendix \section{Calculation of $x(\tau)$}\label{app:xvtau}
For convenience, we introduce reduced dimensionless variables by
$x=R/R_c$ and $\tau=t/t_c$, where $t_c=R_c/\Rpinf$ is the
characteristic time for interface movement. The solution (\ref{tauvx})
of the equation of motion of the bubble radius (\ref{Rpvt}) then takes
the form
\begin{equation} \label{apptauvx}
\tau(x) = x - x_0 + \ln\left(\frac{1+x_0}{1+x}\right),
\end{equation}
which shall now be inverted to yield the bubble radius as a function
of time.

The inverse of $z\mapsto z\,e^z$, is known as Lambert's function $W$
\cite{Lambert}. For real valued $x$, $W$ has two branches:
\begin{eqnarray*}
W_0(x) \geq -1 &;& -\frac{1}{e}\leq x \\
W_{-1}(x)\leq -1 &;& -\frac{1}{e}\leq x \leq 0
\end{eqnarray*}
Exponentiation of (\ref{apptauvx}) yields
\[
-(1+x) e^{-(1+x)}= - (1+x_0)e^{-(1+x_0)} e^{-\tau}.
\]
By definition of $W$ this is equivalent to
\[
x(\tau)=-1 - W_{-1}\left(- (1+x_0)e^{-(1+x_0)} e^{-\tau}\right)
\]
$W_{-1}$ is the right branch, because the argument is always negative,
and $x$ must not be limited.

Note that we reserve $x$ and $\tau$ to denote the reduced bubble
radius and the corresponding reduced time, respectively. Arbitrary
coordinates $(r,t)$ are referred to by $(y,\tet)$, see App.\ B.

\section{Spin wave amplitudes} \label{app:spincalc}
With
$y=r/R_c$, $\tet=t/t_c$, $\varphi(y,\tet)=\Theta_A(r,t)$,
$\vaq=v_A R_c$, $\oaq=\omega_A\,R_c/\csp_A$,  $c_A=\csp_A/\Rpinf$,
Eqs.\ (\ref{sebdyn}) read
\begin{equation} \label{segredwgl}
\vaq=\pdf{y}\,\varphi,\quad \oaq=\pdf{\tet}\,\varphi, \quad
\pdfn{\tet}{2}\varphi-c_A^2 \Delta_y\, \varphi =0,
\end{equation}
while the CoCos
(\ref{sebbc1}) and (\ref{sebbc2}) become
\begin{equation} \label{redsebbc}
0=\Delta \bar{\omega},\quad 0=A \,\pdf{\tau}{x} + \vaq, \quad \mbox{with} \quad
A=-\frac{\Delta\chi}{\chi_A}\frac{\gamma H}{\csp_A^2}\Rpinf\,R_c>0.
\end{equation}
The general solution of (\ref{segredwgl}) is
$\varphi(y,\tet)=F(y-c_A\,\tet)/y$.
Inserting $\vaq\resto{I}=F'\resto{I}/x-F\resto{I}/x^2$ and
$\oaq\resto{I}=-F'\resto{I}/x$, with
\begin{equation} \label{defq}
F\resto{I}=F(x-c_A\tau(x))=q(x),\quad
F'\resto{I}=\frac{\partial_x q(x)}{1-c_A \partial_\tau x},
\end{equation}
into the CoCos
(\ref{redsebbc}), we obtain the following differential equation
for $q(x)$:
\begin{equation} \label{dglq}
\partial_x q(x)=\frac{x (1-c_A)+1}{x(1+x)}q(x)
       - A \left(x\,(1-c_A)+1\right).
\end{equation}
The special solution of (\ref{dglq}) subject to the initial condition
$q(x_0)=0$ is
\begin{eqnarray} \label{ldglq}
&&q(x)=-A\frac{x}{(1+x)^{c_A}} \left(C(x)-C(x_0)\right)\quad \mbox{with} \\
&&C(x)=\frac{1-c_A}{1+c_A}(1+x)^{1+c_A}
+ \frac{1}{c_A}(1+x)^{c_A}
{_2{\rm F}_1}(-c_A,1;-c_A+1;\frac{1}{1+x}). \nonumber
\end{eqnarray}
Here, ${_2F_1}$ denotes the hypergeometric function as defined in
\cite{MW}, Eq.\ (7-76).
Choosing $x_1(y,\tet)$ so that,
\begin{equation} \label{defx1}
F\left(x_1(y,\tet)-c_A\,\tau\left(x_1(y,\tet)\right)\right) = F(y-c_A\,\tet),
\end{equation}
we obtain the spin wave amplitudes outside the bubble as
\begin{eqnarray}
\vaq(y,\tet)&=&\frac{1}{y}\left.\frac{\pdf{x}q(x)}{1-c_A\,\pdf{x}\tau(x)}
 \right|_{x=x_1(y,\tet)} - \frac{1}{y^2}q(x)\resto{{x=x_1(y,\tet)}}
                                               \label{vaqaussen}\\
\oaq(y,\tet)&=&-\frac{1}{y}\left.\frac{\pdf{x}q(x)}{1-c_A\,\pdf{x}\tau(x)}
 \right|_{x=x_1(y,\tet)}.\label{oaqaussen}
\end{eqnarray}
A closed form expression for $x_1(y,\tet)$ is derived in App.\
\ref{app:calcx1}.

\section{Calculation of $x_1(y,\vartheta)$}\label{app:calcx1}

A necessary condition for (\ref{defx1}) to be fulfilled is given by
\[
x_1 -c_A\,\tau(x_1)=y-c_A\,\tet.
\]
Inserting the definition (\ref{tauvx}) of $\tau(x)$ gives after
exponentiation
\[
\frac{1-c_A}{c_A}(1+x_1)\,e^{\frac{1-c_A}{c_A}(1+x_1)}
= \frac{1-c_A}{c_A}(1+x_0)\,e^{\frac{1-c_A}{c_A}}\,
  e^{-x_0}\,
  e^{\frac{y-c_A\,\tet}{c_A}},
\]
so that
\[
x_1(y,\tet)= -1 + \frac{c_A}{1-c_A}W_{-1}\left(
 \frac{1-c_A}{c_A}(1+x_0)\,e^{\frac{1-c_A}{c_A}}\,
  e^{-x_0}\,
  e^{\frac{y-c_A\,\tet}{c_A}}
\right).
\]
To check for the right branch, we make use of the identity
\begin{eqnarray*}
x&=&x_1(x,\tau(x))\\
&=& -1 + \frac{c_A}{1-c_A}W_{-1}\left(
 \frac{1-c_A}{c_A}(1+x)\,e^{\frac{1-c_A}{c_A}(1+x)} \right).
\end{eqnarray*}
Since $c_A >1$, the argument of $W_{-1}$ is negative. $W_{-1}$ is the
right branch, because $x$ has to be a monotonic function of itself.

\newpage
\pagestyle{empty}
\begin{center}
{\huge \bf Figure Captions}
\end{center}
\vspace*{2cm}
{\bf Figure 1:} Mean free path $l_{QP}$ of quasi particles in the
B-phase (taken from Fig.\ 10.1 of Ref.\ \protect\cite{VW}) and
critical bubble radius $R_c$ (as given by Kaul and Kleinert
\protect\cite{Rem1}),both as functions of temperature.
\vspace*{1cm}\\
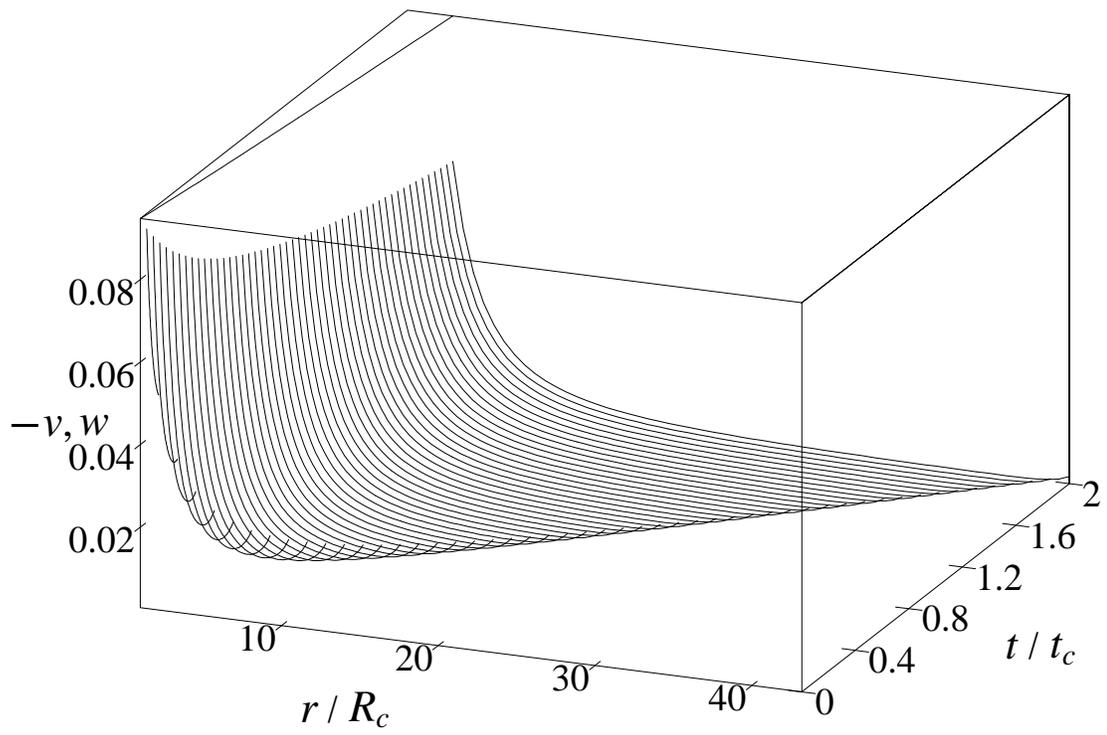
{\bf Figure 2:} Counterflow and spin-superfluid velocity, scaled as\\
$w_A\left(-\frac{\delsig}{\ms}\frac{\rs}{\rho}\Rpinf\right)^{-1}
= -v_A\left(-\frac{\Delta\chi}{\chi_A}\frac{\gamma H}{\csp_A^2}
    \Rpinf\right)^{-1}.$
\vspace*{1cm}\\
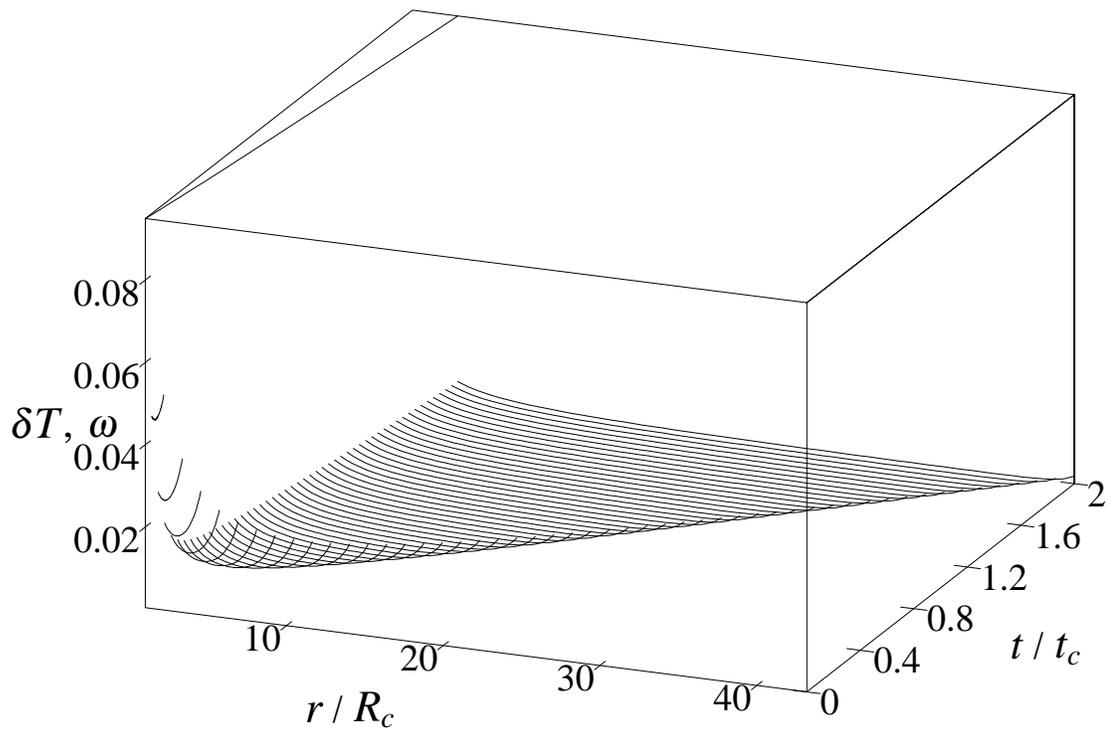
{\bf Figure 3:} Temperature and excess magnetization, scaled as\\
$\dta \frac{\rho\ms}{\rn c_{2A}}
 \left(-\frac{\delsig}{\ms}\frac{\rs}{\rho}\Rpinf\right)^{-1}
= \frac{\omega_A}{\csp_A}
 \left(-\frac{\Delta\chi}{\chi_A}\frac{\gamma H}{\csp_A^2}
    \Rpinf\right)^{-1}.$\tracingstats=1

\begin{thebibliography}{99}
  \bibitem{BSWab} D.~S.~Buchanan, G.~W.~Swift and J.~C.~Wheatley,
   {\it Phys.~Rev.~Lett.}, {\bf 57}, 341 (1986).
  \bibitem{BSmagsig} S.~T.~P.~Boyd and G.~W~Swift,
   {\it Phys.~Rev.~Lett.},{\bf 64}, 894, (1990).
  \bibitem{slowgrowth} S.~T.~P.~Boyd and G.~W~Swift,
   {\it J.~Low.~Temp.~Phys.}, {\bf 87}, 35, (1992).
  \bibitem{fastgrowth} S.~T.~P.~Boyd and G.~W~Swift,
   {\it J.~Low.~Temp.~Phys.}, {\bf 86}, 325, (1992).
  \bibitem{YLabdyn} S.~Yip and A.~J.~Leggett,
   {\it Phys.~Rev.~Lett.}, {\bf 57}, 345 (1986);
   A.~J.~Leggett and S.~K.~Yip, in {\it Helium Three},
   W.~P.~Halperin and L.~P.~Pitaevskii, eds. (Elsevier, Amsterdam 1990);
   A.~J.~Leggett, {\it J.~Low.~Temp.~Phys.}, {\bf 87}, 571, (1992).
  \bibitem{palmeri} J.~Palmeri,
   {\it Phys.~Rev.~Lett.}, {\bf 62}, 1872 (1989);
   N.~B.~Kopnin, {\it Zh.~Eksp.~Theor.~Fiz.}, {\bf 92}, 2106, (1987)
    [{\it Sov.~Phys.~JETP}, {\bf 65}, 1187, (1987)].
  \bibitem{GLab} M.~Grabinski and M.~Liu,
   {\it Phys.~Rev.~Lett.}, {\bf 65}, 2666 (1990).
  \bibitem{panzspinshort} P.~Panzer and M.~Liu,
   {\it Phys.~Rev.~Lett.}, {\bf 69}, 3658, (1992).
   {\it J.~Low.~Temp.~Phys.}, to be published in June 1993
  \bibitem{KoL} P.~Kost\"adt and M.~Liu, {\it to be published.}
  \bibitem{VW} D.~Vollhardt and P.~W\"olfle, {\it The Superfluid
   Phases of Helium 3}, Taylor and Francis, London (1990), sec.~10.2
   and sec.~7.2.
  \bibitem{Rem1}We have put $\alpha=-2\sigma_{AB}$. $\sigma_{AB}$ is the
   interfacial energy as introduced in
   D.~D.~Osheroff and M.~C.~Cross, {\it Phys.~Rev.~Lett.},
   {\bf 38}, 905, (1977);
   R.~Kaul and H.~Kleinert,
   {\it J.~Low.~Temp.~Phys.}, {\bf 38}, 539, (1980);
   N.~Schopohl, {\it Phys.~Rev.~Lett.}, {\bf 58}, 1664, (1987).
  \bibitem{thuneeqbc} E.~V.~Thuneberg,
   {\it Physica B} {\bf 178}, 168 (1992).
  \bibitem{jjmlnonlin} J.~Johannesson and M.~Liu {\it to be published.}
  \bibitem{Lambert} R.~M.~Corless, G.~H.~Gonnet, D.~E.~G.~Hare,
    and D.~J.~Jeffrey, {\it technical report CS-93-03}, Universtiy of
    Waterloo, Canada, (1993). Also available via anonymous ftp from
    {\tt cs-archive.uwaterloo.ca} as {\tt cs-archive/CS-93-03/W.ps.Z}.
  \bibitem{MW} J.~Mathews and R.~L.~Walker, {\it Mathematical Methods
    of Physics}, Addison-Wesley, (1970).
\end{thebibliography}
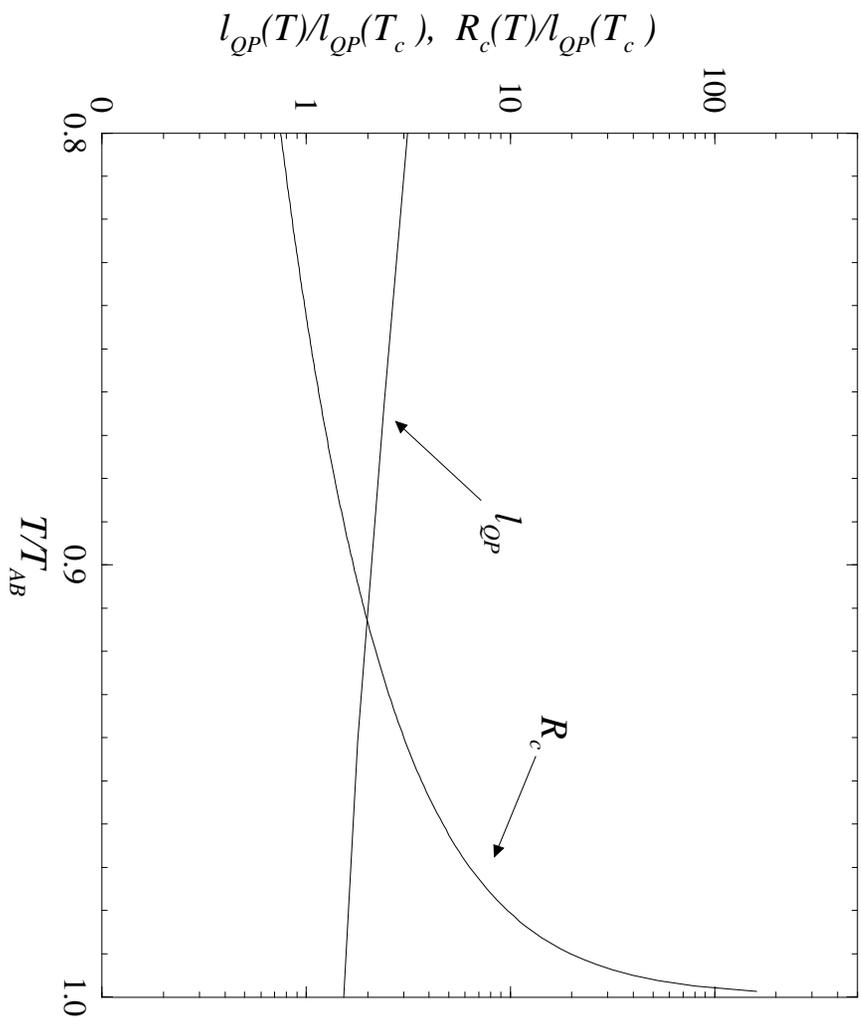
\end{document}